\documentclass[twocolumn,showpacs,preprintnumbers,amsmath,amssymb,prb]{revtex4}
\usepackage{graphicx}
\usepackage{dcolumn}
\usepackage{bm}
\begin{document}
\title{A competing order scenario of two-gap behavior in hole doped cuprates}
\author{Tanmoy  Das,    R.   S.   Markiewicz,   and   A.   Bansil}
\address{Physics  Department,  Northeastern  University,  Boston  MA
02115,                                                          USA}
\date{\today}
\begin{abstract}

Angle-dependent studies of the gap function provide evidence for the
coexistence of two distinct gaps in hole doped cuprates, where the
gap near the nodal direction scales with the superconducting
transition temperature $T_c$, while that in the antinodal direction
scales with the pseudogap temperature. We present model calculations
which show that most of the characteristic features observed in the
recent angle-resolved photoemission spectroscopy (ARPES)  as well as
scanning tunneling microscopy (STM) two-gap studies are consistent
with a scenario in which the pseudogap has a non-superconducting
origin in a competing phase. Our analysis indicates that, near
optimal doping, superconductivity can quench the competing order at
low temperatures, and that some of the key differences observed
between the STM and ARPES results can give insight into the
superlattice symmetry of the competing order.

\end{abstract}
 \maketitle
\narrowtext

\section{Introduction}

The curious angle-dependence of the gap in the hole doped cuprates
has been a subject of intense study for some time now. Initially,
the gap near optimal doping was reported to have the ideal
$|\cos{(k_xa)}-\cos{(k_ya)}|$ form expected for $d-$wave
superconductivity\cite{dwave}, and the deviations observed in
underdoped samples were interpreted as evidence for the presence of
a third-harmonic component in the gap function\cite{mesot}. However,
it has been found recently that the antinodal gap near $(\pi ,0)$
and the nodal gap near $(\pi /2,\pi /2)$ possess distinctly
different doping dependencies in that the antinodal gap follows the
pseudogap, while the nodal gap scales with
$T_c$\cite{tacon,hashimoto,hufner}.  Moreover, in near-optimally
doped Bi$_2$Sr$_2$CaCu$_2$O$_8$ (Bi2212), the gap crosses over from
being of a pure $d-$wave form at low temperatures to one displaying
a pseudogap character above $T_c$\cite{zxshen,kanigel}. This has led
to further debate as to whether the pseudogap is associated with
precursor pairing\cite{renner,eckl,wen,dao} or with a competing
order\cite{deutscher,tacon,hashimoto,hufner,krasnov,tanaka}.  Here
we explore the latter scenario, and show that a model of competing
order can naturally explain a number of puzzling features of both
angle-resolved photoemission spectroscopy
(ARPES)\cite{hashimoto,zxshen,kanigel,tanaka,kondo,terashima,shen}
as well as the more recent scanning tunneling microscopy
(STM)\cite{hoffman,mcelroy,hanaguri,davis} experiments.

We model the pseudogap as a short-range order (SRO) which competes
with $d-$wave superconductivity (dSC) and possesses the symmetry of
the antiferromagnetic (AFM) order of the undoped system. Our
analysis is based mostly on the use of a mean-field
$t-t^{\prime}-t^{\prime\prime}-t^{\prime\prime\prime}-U$ Hubbard
model of competing AFM and dSC orders, and is similar to the one
used previously to successfully describe a number of properties of
the electron-doped cuprates\cite{tanmoy,tanmoyprl}. Such a mean
field treatment has been shown\cite{MKII} to mimic AFM short-range
order, where the Neel temperature $T_N$ approximates the pseudogap
onset temperature $T^*$ and the AFM gap approximates the pseudogap.
Notably, a recent study of the optical properties of
La$_{2-x}$Sr$_x$CuO$_{4}$ (LSCO)\cite{comanac} comes to the
conclusion that the cuprates represent the intermediate coupling
case, which would suggest that the cuprates are amenable to an
approach such as the present one starting from the weak coupling
limit.

Concerning technical details, we include the superconducting (SC)
order empirically through a $d-$wave pairing potential $V$. The
staggered magnetisation $S$ at the nesting vector
$\vec{Q}=(\pi,\pi)$, which gives the pseudogap $US$, as well as the
SC gap $\Delta$ are computed self-consistently at all dopings and
temperatures. The bare dispersion is modeled within a tight-binding
approximation using (in meV): $t=250$, $t^{\prime}=-25$,
$t^{\prime\prime}=12$, and $t^{\prime\prime\prime}=35$. These values
of the hopping parameters are very similar to those adduced earlier
for LSCO\cite{markietb}. Values of Hubbard $U$ and the pairing
potential $V$ have been adjusted to obtain a good fit with the
experimentally observed Fermi surface (FS) arc length and the
overall size of the measured SC gap\cite{uv}.  Although we focus in
this article on the spin density wave (SDW) case, we have also
investigated other ordered phases, including the charge- and
$d$-density wave (C/DDW), and the Pomeranchuk mode.

The paper is organized as follows. Section II discusses SDW-based
results and the corresponding ARPES data on LSCO, while Section III
considers the STM data. Section IV compares predictions of the SDW
model with other candidates for the pseudogap order such as the CDW,
the DDW and the Pomeranchuk mode. In Section V we comment on the
issue of FS arcs vs FS pockets in the light of recent quantum
oscillation experiments. Section VI points out that even though the
magnetic properties of the cuprates are well-known to be quite
asymmetric with respect to electron vs hole doping, the electronic
properties are substantially more electron-hole symmetric. A few
concluding remarks are presented in Section VII. Some of the
relevant technical details of our modeling are given in Appendices
A-C.

\section{Two-gap scenario and the ARPES Data}

\begin{figure}
\rotatebox{270}{\scalebox{0.60}{\includegraphics{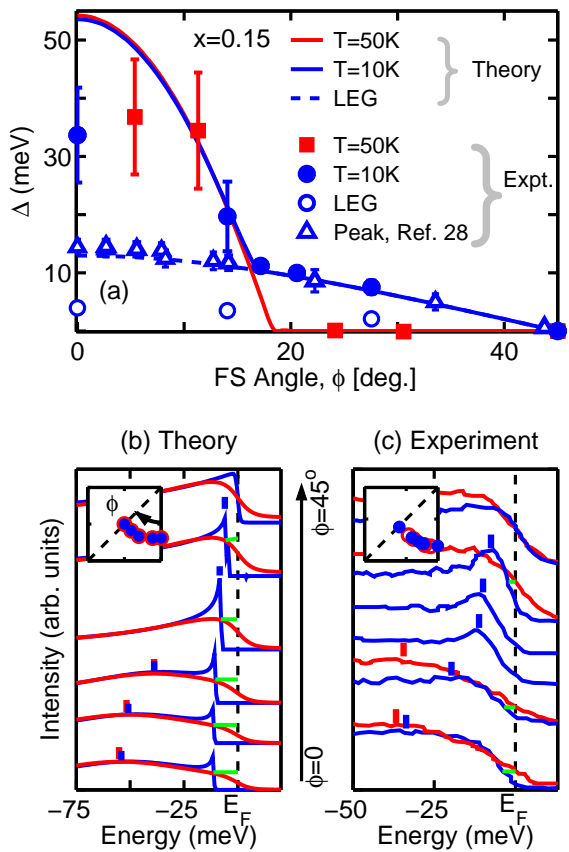}}}
\caption{(color  online)  (a)  Theoretical  and  experimental  angle
dependence  of  various  gaps $\Delta$ in underdoped LSCO ($x$=0.15
) along  the  FS, where $\phi=0$ denotes the antinodal and
$\phi=45^o$ the  nodal  direction.  Solid  lines give our results
for the normal ($T=50$K,  red  line)  and  the superconducting state
($T=10$K, blue line),  and  the  corresponding  experimental  data
are plotted with filled  symbols  of the same color\cite{terashima}.
Blue dashed line shows  the  computed  leading  edge gap (LEG) at
$T=10$K, while blue open circles give the corresponding experimental
LEG as discussed in the  text.  Blue  open  triangles  denote
$\Delta$  based  on  peak positions  from  the  data  of
Ref.~\protect\onlinecite{mesot2}. (b) Computed  energy  distribution
curves  (EDCs) at different momentum points  along  the  FS  ({\it
{see inset}}) for the normal ($T=50$K, blue  lines)  and  the SC
state ($T=10$K, red lines). Spectra at the bottom  of  the  figure
refer to the antinodal direction ($\phi=0$), while  those  at  the
top to the nodal direction ($\phi=45^o$). Blue and  red  tick  marks
on  the spectra denote total gap values while green  lines  mark the
LEG. (c) Same as (b), except that this figure refers  to  the
experimental  EDCs taken from Figs. 2(a) and (b) of
Ref.~\protect\onlinecite{terashima}.  In order to highlight spectral
changes,  normal state spectra (red) are plotted on top of those for
the  SC  state  (blue)  in several cases, even though these pairs of
spectra   are   not   taken  at  exactly  the  same  angle  $\phi$.}
\label{gapangle}
\end{figure}

ARPES experiments in underdoped LSCO report two strikingly different
angle dependencies of the gap,\cite{terashima,mesot2} but find a
natural explanation within our approach. Our calculations for
underdoped LSCO ($x=0.15$) are shown in Fig.~1(a)-(c).  In this
case, our self-consistent  solution  yields  a  staggered
magnetization of $S= 0.21$, which increases very weakly with $T$ up
to $T_c$. The SC gap, i.e.  $\Delta  (T=0)$ is 13meV with $T_c=48$K,
which is higher than the experimental value of 37K\cite{terashima},
reflecting presumably the  neglect  of  phase
fluctuations\cite{emery} in our mean field treatment. Interestingly,
although  the computed ratio $2\Delta/k_BT_c=6.3$ is  anomalously
large, it is in good agreement with experiments\cite{wangdelta}.

We consider first the theoretical results in Figs. 1(a) and (b).
Fig. 1(a) displays calculated and experimental gap values along the
Fermi surface as a function of the angle $\phi$ (inset to
Fig.~1(b)), where $\phi =0$ corresponds to the antinodal direction,
and $\phi =45^o$ to the nodal direction.  Focusing on the energy
distribution curves (EDCs) of the normal state (red spectra)  in
(b), the pseudogap appears as a broad hump feature with a large gap
in the antinodal direction ($\phi=0$) as marked by red tick marks,
which decreases to zero by $\phi=18^o$. For $\phi$ values between
$18^o$ and $45^o$, the FS contains an ungapped nodal pocket or a FS
arc (red curve in (a)). Below $T_c$, the EDCs (blue spectra) show an
additional sharp peak and a leading edge superconducting gap (LEG)
over the whole FS as marked by green lines, while the hump feature
(blue tick marks) remains in the antinodal region. The presence of a
peak-dip-hump feature in the SC state clearly reflects the two gap
behavior, where the peak follows a simple $d-$wave form while the
hump traces the pseudogap. Note that all theoretical spectra in (b)
have been broadened by incorporating the effect of small angle
scattering on the quasiparticles\cite{tanmoy}; see Appendix B for
details. This allows the development of a finite spectral weight at
$E_F$ and the formation of the leading edge gap, even though the
underlying quasiparticle states lie well below $E_F$ at most
momenta\cite{scbroad}.

The  theoretically  predicted  gaps  derived  from the normal and SC
state  spectra  of Fig. 1(b) are plotted in Fig. 1(a), and show good
agreement    with   the   corresponding   experimental   data.   The
characteristics  of the evolution of the theoretical spectra with FS
angle  in  the  presence  of  two gaps, and how these spectra differ
between  the  normal  and  the  SC  state,  as  discussed above  in
connection with Fig. 1(b), are also seen in the experimental
spectra\cite{terashima} of Fig. 1(c). In particular, our theoretical
prediction that the gap is a pure SC gap up to the tip of the FS arc
around $\phi=18^o$ (see (a)), but that it crosses over into becoming
a total gap composed of SC and pseudogap thereafter, gives insight
into various experimental results reported  in  the
literature\cite{terashima,mesot2}.  The peak plotted by Shi {\it et.
al.}\cite{mesot2} corresponds to our calculated SC peak, and as
shown in Fig. 1(a) (blue triangles), this gap  displays a simple
$d-$wave  form.  In  contrast, Terashima {\it et.
al.}\cite{terashima}  consider  the  hump feature and the associated
data  show a two-gap behavior (blue dots in (a)). Nevertheless, in
the  latter  data\cite{terashima} which  is reproduced in (c), the
presence of the $d-$wave LEG can be seen. We have obtained values of
the LEG from the spectra in (c) and plotted these as open circles in
(a).  A  similar  two-gap  behavior can be seen even more clearly in
(Bi,Pb)$_2$(Sr,La)$_2$CuO$_{6+\delta}$ (Bi2201)  data  in Fig.~3 of
Ref.~\onlinecite{kondo}.

\begin{figure}
\rotatebox{270}{\scalebox{0.6}{\includegraphics{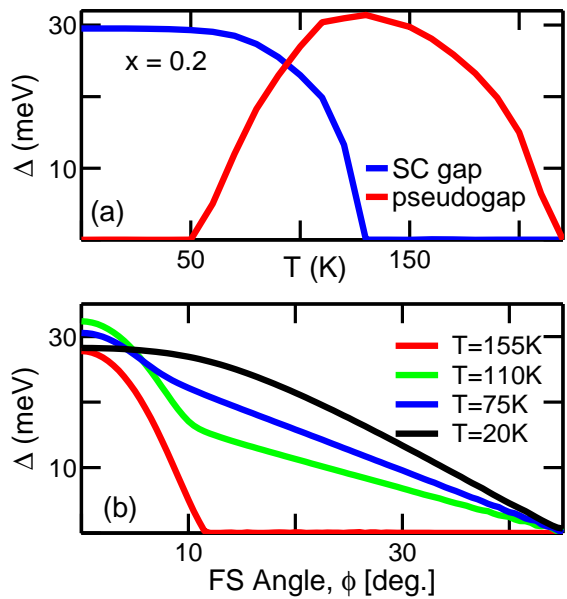}}}
\caption{(color  online) Self-consistently computed results for LSCO
at  $x=0.20$.\cite{uv} (a) Temperature dependence of the SC gap
(blue line)  and  the  pseudogap  (red  line)  showing  reentrant
behavior discussed  in the text. (b) Angle dependence of the gap at
different temperatures.} \label{edc}
\end{figure}

In  the underdoped region, the pseudogap is large compared to the SC
gap.  The  pseudogap  increases  weakly  with  $T$  while the SC gap
decreases  with  $T$.  This  results  in a nearly constant total gap
$\Delta = \sqrt{\Delta_{AFM}^2+ \Delta_{SC}^2}$, which is consistent
with experimental observations\cite{kanigel}. But at higher dopings,
where  the  size  of  the  SC  gap becomes comparable to that of the
pseudogap,  a  more interesting temperature dependence can emerge as
seen  in Fig.~2(a). Here a pseudogap is present at high temperature,
but  after  the SC gap turns on at 130K,\cite{foot3} the pseudogap
is suppressed to  zero  at 50K and thereafter the total gap becomes
a pure SC gap. This  leads  to  the  evolution  of the angle
dependence of $\Delta$ shown  in  Fig.~2(b).  At low temperature
($T=20$K, black line), the gap  is pure $d-$wave, but at high
temperature ($T=155$K, red line), it  is  a  pure  pseudogap, and at
intermediate temperatures the gap shows  a  two-gap  behavior
similar  to  that  of Fig. 1(a). Such a transition  from  a  pure
$d-$wave  pairing gap to a pure pseudogap through  a  region  in
which  both  gaps  coexist has recently been observed  in ARPES
measurements\cite{zxshen,kanigel} on Bi2212 above the  optimal
doping  region. The crossover to pure $d-$wave form at low
temperatures has been taken as evidence that the pseudogap is a
precursor  SC  gap.  In  contrast,  our  analysis  demonstrates that
the appearance of a pure $d-$wave form at low temperatures can
simply be the  replacement  of  one kind of order by a more stable
competitor.

Very different temperature and angle-dependencies in the underdoped
vs optimal/overdoped regions discussed above can be readily
understood.  The pseudogap, which originates here from an ordered
phase, only partially gaps the FS at any finite doping $x$. The SC
gap, on the other hand, opens everywhere, except at the nodal
points, so that if superconductivity is strong enough it can quench
a preexisting ordered pseudogap\cite{bilbro}. While it is natural to
take the competing order to be an SDW in electron-doped cuprates,
where long-range N\'eel order persists up to optimal doping, the
choice is less clear for hole-doping. Accordingly, in Section IV
below, we explore other choices for the pseudogap.

\section{Two Gap Scenario and the STM Data}
\begin{figure}
\rotatebox{270}{\scalebox{0.67}{\includegraphics{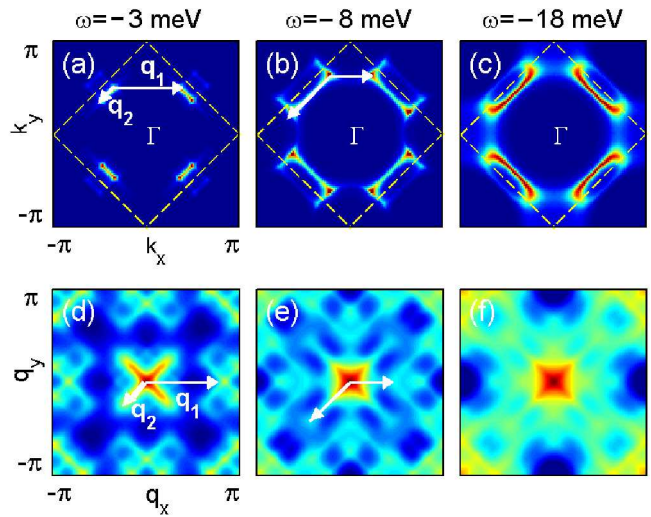}}}
\caption{(color  online)  (a)-(c)  Spectral  intensity  at different
energies $\omega$ below $E_F$. Yellow dashed lines mark the magnetic
zone  boundary. (d)-(f) Corresponding simulated $q-$maps obtained by
convoluting  the  maps  (a)-(c), plotted on a log-scale to highlight
weak  features.  Scattering  vectors  ${\bf{q}}_1$ and ${\bf{q}}_2$,
shown  by  arrows,  are discussed in the text. Red color denotes the
highest   and   blue  the  lowest  intensity.}  \label{stm}
\end{figure}

We turn next to discuss STM results, where also a two-gap behavior
has been reported
recently\cite{hoffman,mcelroy,hanaguri,fujita,davis}.  In STM one
measures the so-called $q-$map, from which the underlying FS and the
angle-dependence of the gap can be extracted by interpreting the
$q-$map as the Fourier transform of a quasiparticle interference
(QPI) pattern\cite{hoffman,mcelroy}.  Figure~3 analyzes the
relationship between $q-$maps and the one-particle spectral weights
within our model, where the spectral weights characterize the ARPES
spectra to the extent that the ARPES matrix element\cite{bansil} can
be neglected.  The computations are based on the LSCO parameters of
Fig.~1 for $x=0.15$ at $T=10$K. In Figs. 3(a)-(c), the computed
spectral weight $A({\bf{k}},\omega)$ is seen to reside mostly in the
momentum region of `bananas' or `arcs' below the AFM zone
boundary\cite{footeh}.  At low $\omega$ the spectral weight is
further concentrated in two bright red spots at the `tips' of each
banana, but at higher energies the weight spreads out more uniformly
over the whole FS arc as seen for example in Fig.~3(c). The
corresponding $q-$maps, modeled as a convolution of the spectral
intensity, i.e. $I_{\bf{q}}=\sum_{\bf{k}}
A({\bf{k}},\omega)A({\bf{k}}+{\bf{q}},\omega)$, are shown in Figs.
3(d)-(f), and display intense peaks at the special $q-$vectors,
which connect the bright spots in
$A({\bf{k}},\omega)$\cite{hanaguri}.  Two such vectors ${\bf{q}}_1$
and ${\bf{q}}_2$ are marked in several panels in Fig.~3 as
examples\cite{foot2}.  It is striking that at high energy in (c),
when the bright tips of bananas in $A({\bf{k}},\omega)$ have
essentially disappeared, the $q-$map in (f) more or less loses its
pattern of well defined peaks as the intensity spreads out over a
wide region.

\begin{figure}
\rotatebox{270}{\scalebox{0.55}{\includegraphics{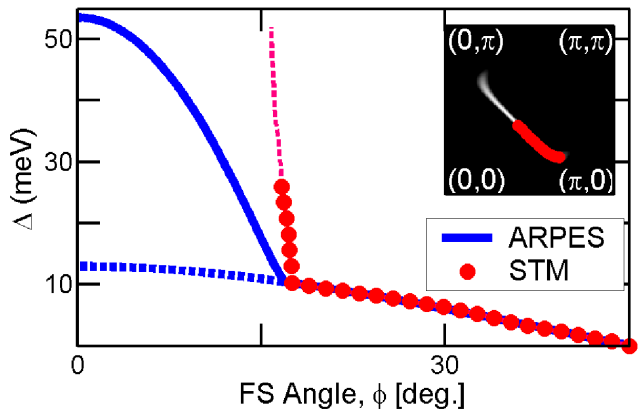}}}
\caption{(color  online)  Apparent gap $\Delta$ as a function of the
FS  angle  (red  dots) as obtained from an analysis of the simulated
$q-$maps  of  Figs. 3(d)-(f). Along the red dashed line the peaks in
the  $q-$maps are quite broad and may be hard to see experimentally.
Blue  solid  line,  which gives the total gap, as well as the dashed
blue  line  giving  the  SC  gap,  is  reproduced from Fig. 1(a) for
reference.  {\it  Inset}: White trace depicts the normal state FS on
which  the FS points obtained from the analysis of $q-$maps of Figs.
3(d)-(f)    are    shown    as    red    dots.   }\label{bsgapangle}
\end{figure}

The simulated $q-$maps of Figs. 3(d)-(f) can be used to reconstruct
the pattern of bright spots in Figs. 3(a)-(c), and to thus obtain
the FS and the angle-dependent gap, as is done commonly in analyzing
STM data. The resulting gap $\Delta$ is seen from Fig. 4 (red dots)
to yield the SC gap in accord with the ARPES data (blue solid line),
but only up to approximately the edge of the FS arc. Interestingly,
for higher $\omega$'s, ARPES follows the pseudogap up to the edge of
the Brillouin zone boundary at $\phi=0$ (blue solid line), but the
STM-derived gaps remain within the AFM boundary up to the end of the
FS arc at $\phi=18^o$. As a result the apparent STM gap adduced from
the $q-$map shoots up nearly vertically at $\phi=18^o$ following the
red dashed line. But with increasing energy above the maximum of the
SC gap of 13meV at $\phi=0$, it becomes difficult to extract
$\Delta$ values as the $q-$map gradually loses its well defined peak
pattern except at $q=0$. Note also that the FS points deduced from
the $q-$maps stop near the AFM zone boundary (see inset to Fig. 4).
These key characteristics of the FS and the angle-dependence of the
gap in Fig. 4 are in remarkable accord with the behavior reported in
a recent STM study of Bi2212\cite{davis}, and reflect the effect of
loss of structure in $q-$maps with increasing energy, which was
pointed out above in connection with Fig.  3.

\section{CDW, DDW and other Orders}
\begin{figure}
\rotatebox{270}{\scalebox{0.55}{\includegraphics{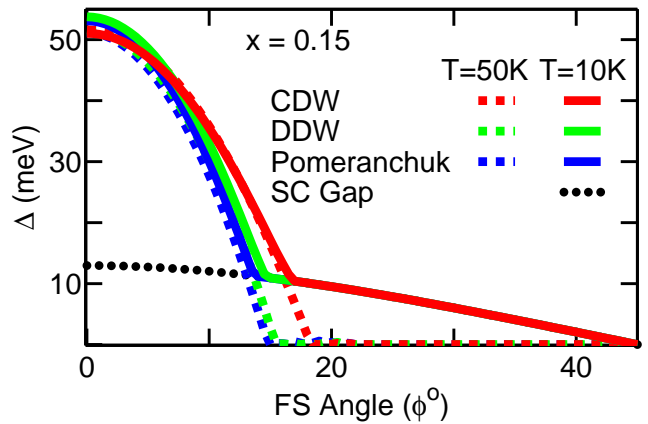}}}
\caption{(color online) Computed angle dependence of the gap
$\Delta$ in the normal (solid lines) and superconducting state
(dashed lines) for various competing orders: CDW (red lines); DDW
(green lines); and Pomeranchuk instability for one axis, which is
taken to be the $y$ axis (blue lines). Black dotted line gives the
pure $d-$wave gap in the absence of any other competing order.}
\label{cdw}
\end{figure}

Although we have focused on properties of the SDW state in this
article, we have also carried out calculations on a number of other
competing electron-hole ordered phases, including the CDW, the DDW,
and the Pomeranchuk mode; see Appendix C for technical details.
Fig.~5, which summarizes our key results, shows that the CDW (red
lines) and DDW (green lines) orders with the same reduced AFM
Brillouin zone as the SDW, yield a gap symmetry very similar to that
of Fig. 1 in the normal as well as the superconducting state. A
doped spin liquid model for the pseudogap\cite{valenzuela} gives
similar results. However, a Pomeranchuk
mode\cite{markiephase,yamase} does not show a true gap, but splits
the van Hove singularity (VHS) along $x$ and $y$ axes. We can,
however, obtain two-gap results similar to those of Fig. 1, along
one axis, as shown by the blue lines in Fig. 5, suggesting that in a
multi-domain sample it might be hard to distinguish this behavior
from the experimental two-gap data.  Finally, we have studied a
linear antiferromagnetic (LAFM) phase\cite{markielafm}, which has a
one-dimensional ordering vector ($\pi,0$), as might be seen in a
stripe phase.  However, we find that the resulting two gap structure
displays a very different symmetry pattern, which is not consistent
with experiments.

\section{Arcs vs Pockets}
\begin{figure}
\rotatebox{270}{\scalebox{0.36}{\includegraphics{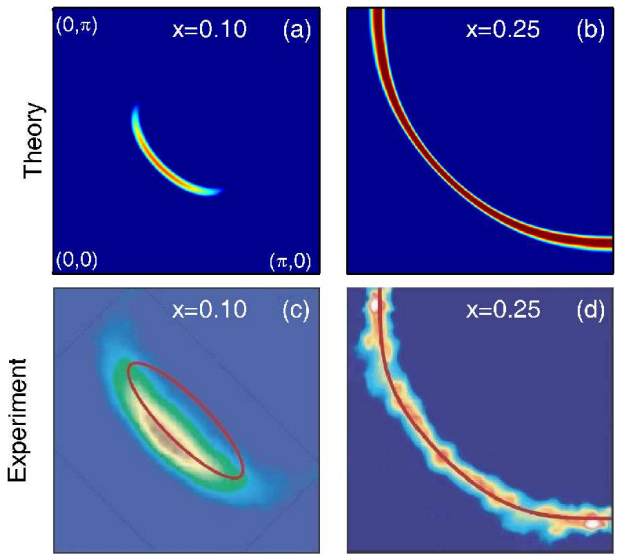}}}
\caption{(color online) (a)-(b) Computed Fermi surface maps for
Na-CCOC based on the AFM model at $x=0.10$ and $x=0.25$. Tight
binding model parameters used in the calculations are: $t=126$,
$t^{\prime}=-36$, $t^{\prime\prime}=10$ meV, with $U=4.5t$ for
$x=0.1$ (gives $S=0.3$) and $U=3.2t$ for $x=0.25$ (gives $S=0$).
(c)-(d) After Ref.~\protect\onlinecite{leyraud}, where the FS's
obtained in Na-CCOC via ARPES from Ref.~\protect\onlinecite{ccoc}
are shown. Red line in (c) depicts the FS of YBCO adduced from
Shubnikov-de Haas experiments of Ref.~\protect\onlinecite{leyraud}
at similar doping levels.} \label{ybco}
\end{figure}

An important issue is whether experiments see a well-defined pocket
or merely a Fermi arc. The recent Shubnikov-de Haas (SdH)
experiments find oscillations in underdoped YBa$_2$Cu$_3$O$_{6.5}$
and YBa$_2$Cu$_4$O$_8$ (YBCO) and argue for the presence of a closed
pocket of approximately the expected
size\cite{leyraud,yelland,bangura}. Note that the AFM model clearly
predicts a full pocket, but the calculated spectral weight resembles
the Fermi arc with little intensity on the shadow side due to the
effect of AFM coherence factors [see Figs.~3(a)-(c) and the inset to
Fig.~4]. Fig.~6 shows results based on the AFM model for another
hole doped cuprate, Ca$_{2-x}$Na$_{x}$CuO$_2$Cl$_2$
(Na-CCOC)\cite{ccoc}. In the underdoped system at $x=0.10$, areas of
the possible FS pocket in Na-CCOC and YBCO would be
similar.\cite{leyraud} While the observed pocket in YBCO (red line
in (c)) is somewhat too small\cite{foot4} to satisfy Luttinger's
theorem, it should be remembered that YBCO has a bilayer splitting
which unlike Bi2212 is not small in the nodal direction. Hence two
nodal FS pockets are expected, where the smaller pocket is
presumably easier to observe in a quantum oscillation
experiment.\cite{foot5} Finally, we note that ARPES data from
underdoped LSCO has recently detected weak spectral weight on the
shadow side of the nodal pocket$-$ see Fig.~2 of
Ref.~\onlinecite{yoshida} for doping $x=0.03$, and $0.07$.
Interestingly, Kaul {\it et al.}\cite{Kaul} point out that this
shadow is consistent with a conventional AFM metal, but it is not
expected in the exotic holon metal phase.

\section{Electron vs hole doping}
\begin{figure}
\rotatebox{270}{\scalebox{0.41}{\includegraphics{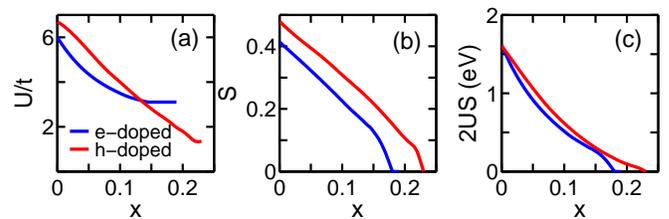}}}
\caption{(color online) Comparison of selected parameters for
electron-doped NCCO/PCCO \cite{tanmoysns} (blue lines) vs hole-doped
LSCO (red lines). (a) Effective interaction $U/t$; (b)
Self-consistent magnetisation $S$; and, (c) Total AFM gap $2US$. }
\label{eh}
\end{figure}

The question of electron-hole symmetry has been a topic of interest
in cuprate physics for some time. The magnetic properties display a
strong asymmetry\cite{damascelli02} in that long-range AFM order
persists up to optimal doping with electron doping, but it
disappears at quite low doping levels in the hole doped case as it
is replaced by the mysterious pseudogap phase. Moreover, nanoscale
phase separation, which is prominent in hole-doped cuprates, seems
to be largely absent in electron doped materials\cite{AAPH}. Some of
these differences could be understood in terms of how the magnetic
susceptibility evolves with electron vs hole
doping.\cite{MKII,water3}

The electronic properties, on the other hand, appear to show greater
symmetry in that superconductivity arises near a possible quantum
critical point (QCP) close to optimal doping, which is associated
with a crossover from small to large Fermi surface
(FS)\cite{nparm,Balak}. In this connection, Fig.~7 compares the
doping dependencies of selected parameters within the AFM model in
electron doped Nd$_{2-x}$Ce$_x$CuO$_2$ (NCCO) and
Pr$_{2-x}$Ce$_x$CuO$_2$ (PCCO) (blue lines) and hole doped LSCO (red
lines). Here, parameters for electron doping are taken from our
earlier work\cite{tanmoysns}. The effective $U$ in Fig.~7(a)
decreases almost linearly with doping in the very underdoped region
in a very similar manner for both electron and hole doping.  At
higher doping, the smaller $U$ found for hole doping could be
associated with larger screening resulting from the proximity of the
VHS. The self-consistently calculated value of the staggered
magnetization $S$, however, remains higher than that on the electron
doped side over the entire doping range as seen in Fig.~7(b).
Remarkably, the total AFM gap $2US$ ($t=0.326$~eV for electron
doping) displays electron-hole symmetry, although the QCP is
slightly higher for hole doping.

The effect of the VHS is reflected in the superconducting properties
as well. At $x=0.15$, for example, the self-consistently calculated
values of the superconducting order parameters are as follows: for
electron (hole) doping, $V=-134 (-78.4)$ meV and $\Delta=2.6 (13)$
meV. That is, even though the interaction $V$ is much weaker for the
hole-doped case, the SC gap is found to be larger, leading to a
larger ratio $2\Delta/k_BT_c$ of 6.3 for hole doping compared to 4.1
for electron doping.

In the electron-doped cuprates the QCP involves two topological
transitions as a function of doping.  At low doping, the FS consists
of electron pockets near $\Gamma$.  As doping increases, a second
hole-like pocket emerges along the nodal direction around 15\%
doping (in NCCO) as the magnetic gap decreases and the lower
magnetic band crosses the Fermi level.  The appearance of this
pocket can be detected in ARPES, Hall effect, and penetration depth
measurements at essentially the same doping level\cite{tanmoysns}.
At higher doping, the gap collapses and the FS evolves into a single
large hole-like sheet.  A reverse scenario seems to be followed for
hole doping: A nodal hole-like FS pocket is present at the lowest
dopings (corresponding to the Fermi arc), while at higher dopings
Hall effect evidence has been found\cite{Balak} for the appearance
of electron-like pockets in Bi2201, again at approximately 15\%
doping.  In our model, these would be the $\Gamma$-centered electron
pockets associated with the upper magnetic band.

\section{Conclusion}

In conclusion, our model calculations show that most of the
characteristic features observed in ARPES as well as STM two-gap
studies are consistent with a scenario in which the pseudogap has a
non-superconducting origin in a competing phase. In contrast, a
precursor superconductor model of the pseudogap will have
difficulties explaining why the higher order harmonic content of the
gap grows with increasing temperature as seen in Fig. 2. Our
computed spectra not only show the presence of a feature which
scales with the pseudogap, but also display a superconducting
low-energy gap leading to a peak-dip-hump structure. As doping
increases toward the quantum critical point of the pseudogap, we
find that a region of reentrant superlattice order could reappear in
the system. Our analysis highlights electron-hole symmetry of
electronic properties of the cuprates, even though the magnetic
properties are well-known to be quite asymmetric.

\begin{acknowledgments}

This work is supported by the U.S.D.O.E contracts DE-FG02-07ER46352
and DE-AC03-76SF00098 and benefited from the allocation of
supercomputer time at NERSC and Northeastern University's Advanced
Scientific Computation Center (ASCC).

\end{acknowledgments}
\appendix
\addcontentsline{toc}{section}{Appendix}
\section{Model for the coexistence of SDW and $d-$SC orders}
Models of competing SDW and SC order have been studied for many
years\cite{machida,gabovich,bilbro}. In our calculations, we use a
one band tight binding model Hamiltonian where antiferromagnetism is
included via a Hubbard $U$ and superconductivity as in the BCS
theory:\cite{tanmoy}
\begin{eqnarray}\label{h}
H &=&
\sum_{\vec{k},\sigma}\xi_{\vec{k}}c_{\vec{k},\sigma}^{\dag}c_{\vec{k},\sigma
} +
U\sum_{\vec{k},\vec{k^{\prime}}}c_{\vec{k}+\vec{Q},\uparrow}^{\dag}c_{\vec{k
},\uparrow}
c_{\vec{k^{\prime}}-\vec{Q},\downarrow}^{\dag}c_{\vec{k}^{\prime},\downarrow}\nonumber\\
&&+\sum_{\vec{k},\vec{k^{\prime}}}V(\vec{k},\vec{k^{\prime}})
c_{\vec{k},\uparrow}^{\dag}c_{-\vec{k},\downarrow}^{\dag}
c_{-\vec{k^{\prime}},\downarrow}^{\dag}c_{\vec{k}^{\prime},\uparrow}^{\dag},
\end{eqnarray}
where $c_{\vec{k},\sigma}^{\dag} (c_{\vec{k},\sigma})$ is the
electronic creation (destruction) operator with momentum $\vec{k}$
and spin $\sigma$ ($\bar{\sigma}$ is the opposite spin). The bare
particle dispersion with respect to the chemical potential $E_F$ is
given by
\begin{eqnarray}\label{xi}
&&\xi_{\vec{k}}=-2t[c_x(a)+c_y(a)]
-4t^{\prime}c_x(a)c_y(a)\nonumber\\
&&~~~~~~-2t^{\prime\prime}[c_x(2a)+c_y(2a)]\nonumber\\
&&~~~~~~-4t^{\prime\prime\prime}[c_x(2a)c_y(a)+c_x(a)c_y(2a)]-E_F,
\end{eqnarray}
with $c_i(\alpha a)=cos(\alpha k_i a)$ and $a$ being the lattice
constant. $t, t^{\prime},t^{\prime\prime}$ and
$t^{\prime\prime\prime}$ are TB hopping parameters. Defining the
Nambu operator in the magnetic Brillouin zone (MBZ)
\begin{equation}\label{psi}
\Psi_{\vec{k}}=\left(\begin{array}{c}
c_{\vec{k},\uparrow}\\~~~c_{\vec{k}+\vec{Q},\uparrow}\\~c_{-\vec{k},
\downarrow}^{\dag}\\~~~~c_{-\vec{k}-\vec{Q},\downarrow}^{\dag}
\end{array}
\right)
\end{equation}
we can write the above Hamiltonian in the MBZ as
\begin{equation}\label{hm}
H =\left(\begin{array}{ccccc}
~\xi_{\vec{k}} & -US &~ \Delta_{\vec{k}} &~0\\
-US &~\xi_{\vec{k}+\vec{Q}}&~0&~~\Delta_{\vec{k}+\vec{Q}}\\
 ~\Delta_{\vec{k}}&0&-\xi_{\vec{k}}&-US\\
~0 & ~\Delta_{\vec{k}+\vec{Q}}&-US &~-\xi_{\vec{k}+\vec{Q}}
\end{array}
\right)
\end{equation}
where the order parameters $S$ and $\Delta_{\vec{k}}$ represent the
staggered magnetization at nesting vector $\vec{Q}=(\pi,\pi)$ and
the superconducting gap, respectively. In the mean field
approximation, these are defined by
\begin{eqnarray}\label{or}
S&=&\sum_{\vec{k},\sigma}\sigma\left\langle
c_{\vec{k}+\vec{Q},\sigma}^{\dag}c_{\vec{k},\sigma}\right\rangle,\\
\Delta_{\vec{k}}&=&\sum_{\vec{k^{\prime}}}V(\vec{k},\vec{k^{\prime}})
\left\langle
c_{\vec{k^{\prime}},\uparrow}^{\dag}c_{-\vec{k^{\prime}},\downarrow}^{\dag}
\right\rangle=\Delta g_{\vec{k}}\nonumber\\
&=&Vg_{\vec{k}}\sum_{\vec{k^{\prime}}}g_{\vec{k^{\prime}}}\left\langle
c_{\vec{k^{\prime}},\uparrow}^{\dag}c_{-\vec{k^{\prime}},\downarrow}^{\dag}
\right\rangle=-\Delta_{\vec{k}+\vec{Q}},
\end{eqnarray}
where the $d_{x^2-y^2}-$orbital phase factor is
$g_{\vec{k}}=[c_x(a)-c_y(a)]/2$. We diagonalize the Hamiltonian of
Eq. \ref{hm} by the Bogolyubov method and the corresponding unitary
matrix can be easily constructed\cite{zaira},
\begin{equation}
\hat{U}_{\vec{k},\sigma}=\left(\begin{array}{cccc}
\alpha_{\vec{k}}u^{+}_{\vec{k}}
&\sigma\beta_{\vec{k}}u_{\vec{k}}^{-} & -\alpha_{\vec{k}}
v_{\vec{k}}^{+}
& \sigma\beta_{\vec{k}}v^{-}_{\vec{k}}\\
-\sigma \beta_{\vec{k}}u^{+}_{\vec{k}}
&\alpha_{\vec{k}}u_{\vec{k}}^{-} &
\sigma\beta_{\vec{k}}v_{\vec{k}}^{+} &
\alpha_{\vec{k}}v^{-}_{\vec{k}}\\
\alpha_{\vec{k}}v^{+}_{\vec{k}} &
-\bar{\sigma}\beta_{\vec{k}}v_{\vec{k}}^{-} &
\alpha_{\vec{k}}u_{\vec{k}}^{+} &
\bar{\sigma}\beta_{\vec{k}} u^{-}_{\vec{k}}\\
-\bar{\sigma}\beta_{\vec{k}}v^{+}_{\vec{k}} &
-\alpha_{\vec{k}}v_{\vec{k}}^{-} &
-\bar{\sigma}\beta_{\vec{k}}u_{\vec{k}}^{+} &
\alpha_{\vec{k}}u^{-}_{\vec{k}}\\\end{array} \right).
\end{equation}
The Bogolyubov coefficients are chosen to be
\begin{eqnarray}\label{cf}
\alpha_{\vec{k}}
(\beta_{\vec{k}})&=&\frac{1}{2}\sqrt{1\pm\frac{\xi_{\vec{k}}^-}{E_{0\vec{k}}
}}\\
u_{\vec{k}}^{\nu}
(v_{\vec{k}}^{\nu})&=&\frac{1}{2}\sqrt{1\pm\frac{\xi_{\vec{k}}^+ +
\nu E_{0\vec{k}}}{E^{\nu}_{\vec{k}}}},
\end{eqnarray}
where
$\xi_{\vec{k}}^{\pm}=(\xi_{\vec{k}}\pm\xi_{\vec{k}+\vec{Q}})/2$ and
$E_{0\vec{k}}=\sqrt{(\xi_{\vec{k}}^-)^2+(US)^2}$.  The four
resulting quasiparticle bands have energies $\pm E^{+}_{\vec{k}}$,
$\pm E^{-}_{\vec{k}}$, where
\begin{equation}\label{band}
(E^{\nu}_{\vec{k}})^2=(\xi_{\vec{k}}^++\nu E_{0\vec{k}})^2 +
\Delta_{\vec{k}}^2,
\end{equation}
and $\nu=\pm$ refers to the upper (+) and lower (-) magnetic bands
(U/LMB). The $4\times4$ Matsubara Green's function can be defined
from the Nambu operator as
$G(\vec{k},\tau-\tau^{\prime})=-\left\langle
T_{\tau}\Psi_{\vec{k}}(\tau)\Psi_{\vec{k}}^{\dag}(\tau^{\prime})
\right\rangle$, whose Fourier transformation gives,
\begin{equation}\label{gn}
G(\vec{k},\sigma,i\omega_n)=\hat{U}_{\vec{k},\sigma}(i\omega_n-H_{\rm{diag}}
)^{-1}\hat{U}_{\vec{k},\sigma}^{\dag}
\end{equation}
where $H_{\rm{diag}}$ is the diagonalized Hamiltonian containing the
eigenvalues in order $[E_{\vec{k}}^+, E_{\vec{k}}^-, -E_{\vec{k}}^+,
-E_{\vec{k}}^-]$. The corresponding $4\times4$ spectral function is
defined in the standard form
$A(\vec{k},\sigma,i\omega_n)=-{\rm{Im}}[G(\vec{k},\sigma,i\omega_n)]/\pi$.
We calculate the self-consistent values of various order parameters
for each hole doping $x$, by simultaneously solving the following
set of equations
\begin{eqnarray}\label{sc}
x &=& 1
-\sum_{\vec{k},\sigma}\int_{-\infty}^{\infty}\frac{d\omega}{2\pi}A_{11}(\vec
{k},\sigma,\omega+i\delta)f(\omega),\\
S &=&
\sum_{\vec{k},\sigma}\sigma\int_{-\infty}^{\infty}\frac{d\omega}{2\pi}A_{12}
(\vec{k},\sigma,\omega+i\delta)f(\omega),\\
{\rm{and}}\nonumber\\
\Delta &=&
V\sum_{\vec{k},\sigma}\int_{-\infty}^{\infty}\frac{d\omega}{2\pi}A_{13}(\vec
{k},\sigma,\omega+i\delta)f(\omega).
\end{eqnarray}
$f(\omega)=1/(\exp{(\omega/k_BT)}+1)$ is the Fermi function at
temperature $T$ where $k_B$ is the Boltzmann constant.

\section{Broadening due to small angle scattering}
In both ARPES and STM calculations we model the large spectral
broadening to be due to elastic small angle scattering of the Cooper
pairs (neglecting pair breaking effects), in which case the Green's
function remains of the same form but with renormalized parameters
$\omega \rightarrow \tilde{\omega} = \omega Z_{\vec{k}}(\omega)
 $ and
$\Delta_{\vec{k}}\rightarrow\tilde{\Delta}_{\vec{k}}=\Delta_{\vec{k}}Z_{\vec
{k}}(\omega)$. The renormalization factor is taken to
be\cite{foot1},
\begin{equation}\label{broad}
Z_{\vec{k}}(\omega)=1+\frac{i\Sigma_{\vec{k}}(\omega)}{\sqrt{\omega^2-\Delta
_{\vec{k}}^2}},
\end{equation}
where $\Sigma_{\vec{k}}(\omega)$ is the self energy correction due
to impurity scattering, which is related to the normal state
scattering rate as\cite{tanmoy}
\begin{equation}\label{broad1}
\Sigma_{\vec{k}}(\omega) = {\rm{sgn}}(\omega)(C_0+C_1\omega^p).
\end{equation}
The power $p=3/2$ is assumed to apply for holes, as found for
electron doped cuprates\cite{tanmoy}. The other parameters $C_0$ and
$C_1$ are found by fitting to the experimental broadening. For ARPES
calculations $C_0=C_1=50$meV. For STM, the $\omega-$dependence is
neglected and $C_0=30$meV.
\section{Other models of competing order}

We have also studied various other ordered phases such as the CDW,
the DDW, the Pomeranchuk instability, and the linear
antiferromagnetism (LAFM). All these orders possess the same
superlattice behavior and/or the pseudospin character near the Van
Hove singularity and thus are possible candidates for the origin of
the pseudogap. When the Hamiltonian is solved in the mean field
approximation, the eigenvalues are similar to Eq. \ref{band} for SDW
except that $E_{0\vec{k}}$ is different for each
phase.\cite{markiephase,markielafm} Defining the interaction as
$V_i$ with $i=\rm{CDW,DDW,Pom,LAFM}$, $E_0$ can be written as
\begin{equation}\label{allphase}
E_{0\vec{k}}=\sqrt{R_x^2+R_y^2+R_{z,\sigma}^2 +
(\xi_{\vec{k}}^-)^2+2\xi_{\vec{k}}^-R_x},
\end{equation}
where $R_i$ represent the gap in various
phases\cite{markiephase,markielafm}; in Pomeranchuk mode
$R_x=V_{\rm{Pom}}g_{\vec{k}}$, in DDW phase $R_y =
V_{\rm{DDW}}g_{\vec{k}}$. In CDW and LAFM phase the gap is similar
to the SDW phase except that in CDW the gap does not depend on the
spin orientation of the system and thus $R_{z,\sigma}=V_{\rm{CDW}}$.
The LAFM leads to a similar result, with $R_{z,\sigma}=\sigma
V_{\rm{LAFM}}S$, except here the nesting of the FS occurs along
$\vec{Q}=(\pi,0)$. In each case the interaction was adjusted to
match the gap and the FS arc for LSCO at $x=0.15$ [see Fig.~5] The
resulting values are $V_{\rm{DDW}}=200$, $V_{\rm{CDW}}=160$ meV,
whereas in the Pomeranchuk mode, $R_x$ simply renormalizes $t$ to
have unequal values of $x$ and $y$, $t_{x/y}=t\pm1.2$ meV. We find
that LAFM has the wrong pseudogap symmetry to explain the
experiments.

\begin{thebibliography}{99}
\bibitem{uv}
The values of the interaction potentials are $U= 2.8t$ and $V=-78.4$
meV at $x=0.15$.  For $x=0.2$, the parameters are chosen to
illustrate the reentrance of the pseudogap as described in the text:
$U=1.5t$ and $V=-51$meV.
\bibitem{scbroad}The  SC  gap,  however,  has  not  been  broadened.
\bibitem{footeh} We find a small electron-hole asymmetry in the
$q$-maps.
\bibitem{foot4}In an AFM model, the observed oscillation should
correspond to two pockets associated with inequivalent nodal points.
\end{thebibliography}
\end{document}